\documentclass[9pt,twocolumn,twoside]{osajnl}

\journal{ol} 

\setboolean{shortarticle}{true}

\title{Optimal output coupler grating reflectivity for Er/Yb fiber lasers}

\author[1,*]{Z. Montz}
\author[1,2]{A. Shirakov}
\author[1]{U. Ben Ami}
\author[1]{S. Genish}
\author[1]{A. A. Ishaaya}

\affil[1]{Department of Electrical and Computer Engineering, Ben-Gurion University of the Negev, Beer Sheva, 8410501, Israel}
\affil[2]{Department of Physics, Ben-Gurion University of the Negev, Beer Sheva, 8410501, Israel}

\affil[*]{Corresponding author: zevm@post.bgu.ac.il}

\ociscodes{(060.2310) Fiber optics; (060.3510) Lasers, fiber; (060.3735) Fiber Bragg gratings; (060.7140) Ultrafast processes in fibers.}

\usepackage{amsmath,amssymb,graphicx,physics,caption,gensymb}
\usepackage{textcomp} 
\usepackage{cleveref} 
\usepackage{siunitx} 

\crefformat{equation}{(#2#1#3)} 

\hypersetup{
  colorlinks   = true, 
  urlcolor     = blue, 
  linkcolor    = blue, 
  citecolor    = blue  
}

\begin{abstract}
We measure the output power of an Er/Yb fiber laser with twelve different~\mbox{SMF-28} narrowband output couplers and demonstrate experimentally that the optimal reflectivity is~${\sim1~\%}$. The fiber laser efficiency with the optimal output coupler is~${\sim38~\%}$. In addition, we successfully inscribe a similar output coupler in-situ during laser operation with~${800~nm}$ femtosecond pulses and the phase mask technique. An output power very close to the optimal was obtained with the in-situ inscribed output coupler.
\end{abstract}

\setboolean{displaycopyright}{true}

\begin{document}

\setlength{\parskip}{0pt} 
\setlength{\textfloatsep}{10pt}

\setlength\abovedisplayskip{-5pt}
\setlength\belowdisplayskip{5pt}
\setlength\abovedisplayshortskip{-5pt}
\setlength\belowdisplayshortskip{5pt}

\maketitle

\noindent
Fiber Bragg grating (FBG) fabrication methods have developed rapidly after the first theoretical foundations of periodic stratified media~\cite{ash1970grating,yeh1977electromagnetic}. Permanent FBGs were first demonstrated by illuminating a photosensitive fiber core with intense visible counter-propagating coherent beams~\cite{hill1978photosensitivity}. This method quickly evolved to a more convenient and versatile side illumination method with two coherent UV beams~\cite{meltz1989formation}. Several years later, FBGs were inscribed by placing a phase mask between the fiber and the cylindrical lens with UV laser sources~\cite{anderson1993production,hill1993bragg}. Recently, FBGs were inscribed with the point-by-point~\cite{kondo1999fabrication}, plane-by-plane~\cite{theodosiou2016modified} and phase mask~\cite{mihailov2003fiber} techniques using various femtosecond laser sources. The phase mask technique seems more suitable for mass production due to the high precision FBGs inscribed with the phase mask interference pattern. Compared with the phase mask technique, the point-by-point and plane-by-plane techniques are more versatile and dynamic; yet, the inscribed Bragg wavelength is limited by the focused laser spot size (difficult to inscribe FBGs with short periods). FBGs inscribed with femtosecond lasers have shown to have a better temperature resistance than FBGs inscribed with standard UV laser sources such as the KrF excimer laser and the second harmonic of an Ar-ion laser~\cite{mihailov2004bragg}. The permanent refractive index induced with ultra-short pulses is caused by a nonlinear multiphoton absorption mechanism, thus allowing to inscribe FBGs on non-photosensitive fibers. Improvement in FBG fabrication methods mentioned above ultimately led to wide-scale integration of FBGs in fiber lasers.

Fiber lasers typically have two FBGs inscribed on passively matched double-clad fibers that are spliced to the active fiber. The first FBG is a high reflecting mirror and the second FBG, also known as the output coupler~(OC), is a partially reflecting mirror. Every laser has a precise OC that optimizes its output power for a given pump power. An approximation of the optimal OC reflection can be calculated with specified laser parameters such as internal cavity losses, gain, saturation intensity and pump power~\cite{hodgson2005laser}. Yet, the exact values of these parameters are usually not known; therefore, the OC reflection is typically determined experimentally by trial and error. In order to find the most suitable OC it is common to choose several OCs, splice each one to the active fiber and measure the output power at the desired pump power. In practice, most fiber laser manufacturers use OCs with reflections higher than~${\sim4~\%}$, which is not optimal in many cases.

Here, we measure the optimal narrowband OC grating reflectivity of an Er/Yb fiber laser with two FBGs in the resonator, and compare the results to a homogeneously broadened continuous-wave~(CW) laser model with uniform distributed loss. The model is in good agreement with our measurements for OCs with reflectivities higher than~${\sim3~\%}$; yet, the model fails to predict the optimal OC reflectivity. Interestingly, the optimal OC reflection for a~${\sim18.27~W}$ pump power is much lower than typical reflection values used in commercial Er/Yb fiber lasers.

\begin{figure}[!t]
\centering
\includegraphics[width=\linewidth]{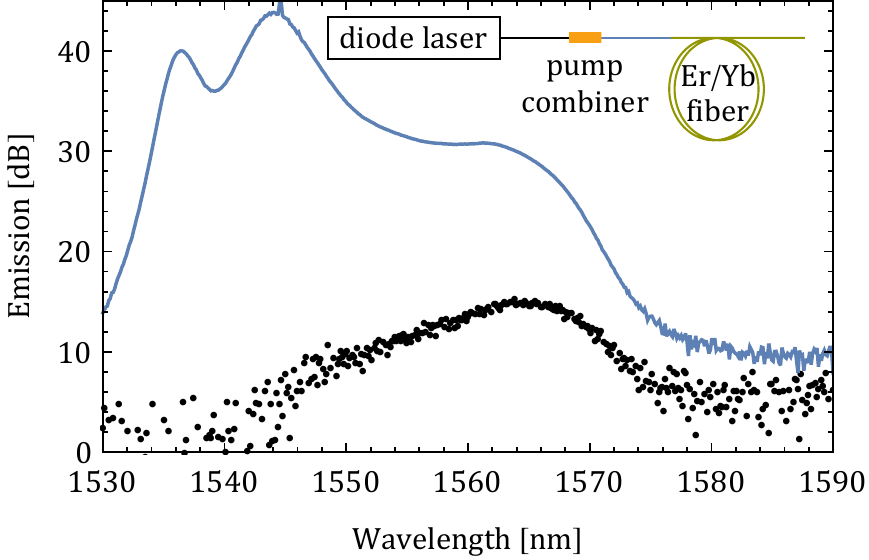}
\captionsetup{justification=justified}
\caption{Emission spectrum of the Nufern~\mbox{MM-EYDF-10/125} double-clad fiber for pump powers of~${1.4~W}$ (black circles) and~${3.22~W}$ (blue curve). Figure inset shows the emission measurement layout.}
\label{fig:1}
\end{figure}

Prior to optimizing the OC reflectivity we measured the emission spectrum of the active double-clad fiber in order to approximate the optimal center wavelength of the fiber laser. In~Fig.~\hyperref[fig:1]{\ref*{fig:1}} we show the emission spectrum of the active fiber with two different pump powers. The active fiber of the home-made fiber laser was a~${\sim4~m}$ Nufern~\mbox{MM-EYDF-10/125} double-clad fiber. Diode pump laser center wavelength and maximum output power were~${915~nm}$ and~${60~W}$, respectively. The spectrum was monitored with an Anritsu MS9710B optical spectrum analyzer~(OSA). At very low pump powers the highest emission was at~${1563.8~nm}$; yet, as the pump power increased the highest emission blue shifted towards~${1544.6~nm}$. Pump powers corresponding to these wavelengths were~${1.4~W}$ and~${3.22~W}$, respectively. Following the active fiber emission measurements, we inscribed a high reflecting mirror on a passive double-clad single-mode fiber~(${1550~nm}$ matched) with a center wavelength near the highest emission of the active fiber. The inscribed high reflecting mirror had the following specifications: wavelength bandwidth~${1\pm0.1~nm}$, center wavelength~${\sim1545.6~nm}$, and reflectance~${>99.5~\%}$.

The setup for inscribing the OCs is presented in the inset of~Fig.~\hyperref[fig:2]{\ref*{fig:2}(a)}. A cylindrical plano-convex lens with a~${30~mm}$ focal length was used to inscribe the OCs on stripped Corning~\mbox{SMF-28} fibers. We fine-tuned the inscribed wavelength by placing a cylindrical plano-convex lens with a~${1000~mm}$ focal length~${\sim0.4~m}$ before the~${30~mm}$ cylindrical lens. Shifting the inscribed Bragg wavelength by placing an additional cylindrical lens in the beam path is similar to the method proposed in~\cite{prohaska1993magnification}. The plano-convex cylindrical lens converges the wavefront and deceases the inscribed Bragg wavelength. The Coherent Legend Elite femtosecond laser was operating at a center wavelength of~${800~nm}$ with pulse duration and repetition rate of~${\sim40~fs}$ and~${1~kHz}$, respectively. Average power to inscribe the low reflecting OCs~${<2~\%}$ and high reflecting OCs~${>2~\%}$ was~${60-80~mW}$ and~${80-110~mW}$, respectively. The phase mask (Ibsen Ltd.) was designed for first-order inscription with a~${1070~nm}$ period. The stripped fiber was placed~${\sim2~mm}$ away from the phase mask during the inscription to ensure pure two-beam interference~\cite{smelser2004generation}. OC transmission and reflection measurements were performed with a Yokogawa AQ6370D OSA. The relationship between the OC reflectivity~${R}$ and difference between the transmitted ASE and signal (in~${dB}$) when measuring the mirror in transmission is expressed with the following equation:

\begin{figure}[!t]
\centering
\includegraphics[width=\linewidth]{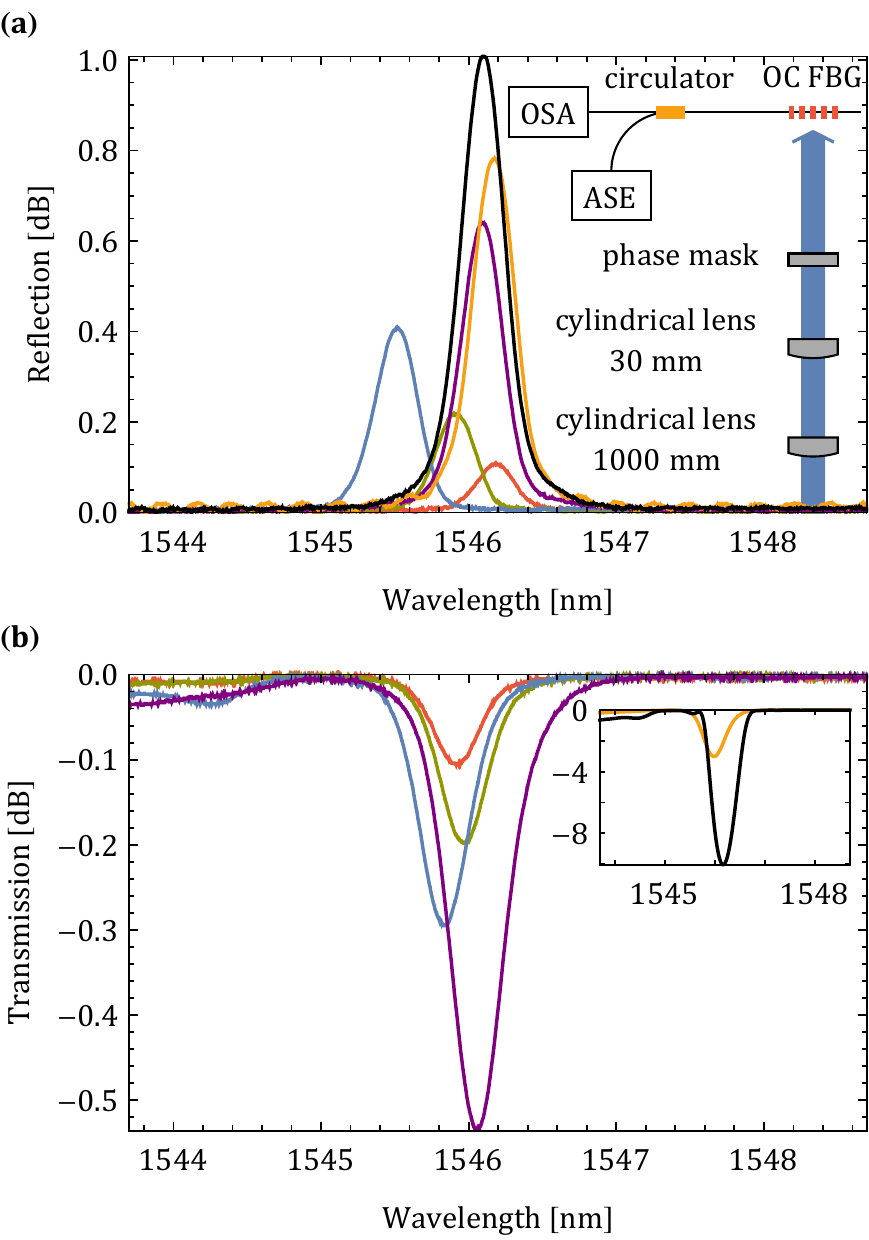}
\captionsetup{justification=justified}
\caption{(a) Reflection measurements of the low reflecting OCs~${<2~\%}$ and (b) transmission measurements of the high reflecting OCs~${>2~\%}$. ASE signal was subtracted from all measurements. Figure inset (a) shows the inscription setup and figure inset (b) shows the transmission measurement of the OCs with highest reflectivity.}
\label{fig:2}
\end{figure}

\begin{equation}
\label{eq:1}
\Delta_{t} = 10\log_{10}\left(1-R\right)
\end{equation}

\noindent
where~${\Delta_{t}}$ is the measured difference between the ASE transmission without an inscribed OC and signal minimum transmission with an OC. The resolution of the OSA~${0.1~dB}$ corresponds to a reflectivity of~${\sim2.28~\%}$ when measuring the mirror in transmission. In order to check the performance of the fiber laser with low reflecting OCs~${<2~\%}$ we inscribed OCs and monitored their reflectivity. The relationship between the OC reflectivity~${R}$ and difference between the reflected ASE and signal when measuring the mirror in reflection is expressed with the following equation~\cite{montz2019inscribing}:

\begin{equation}
\label{eq:2}
\Delta_{r} = 10\log_{10}\left(\dfrac{R+(1-R)^{2}F/(1-RF)}{F}\right)
\end{equation}

\noindent
where~${\Delta_{r}}$ is the measured difference between the ASE flat cleaved Fresnel reflection without an inscribed OC and signal maximum reflection from a flat cleaved OC. ${F}$ is the calculated flat cleaved Fresnel reflection. An approximation of the Fresnel reflection from the flat cleaved fiber facet was calculated with the effective index of an~\mbox{SMF-28} at~${1550~nm}$, which is~${\sim1.4682}$. The resolution of the OSA corresponds to a reflectivity of~${\sim0.09~\%}$ when measuring the mirror in reflection. OC transmission and reflection measurements are shown in~Fig.~\hyperref[fig:2]{\ref*{fig:2}}. The reflections of the inscribed OCs in an increasing order were:~${\sim0.10~\%}$, ${\sim0.20~\%}$, ${\sim0.38~\%}$, ${\sim0.61~\%}$, ${\sim0.76~\%}$, ${\sim1.01~\%}$, ${\sim2.41~\%}$, ${\sim4.46~\%}$, ${\sim6.57~\%}$, ${\sim11.61~\%}$, ${\sim49.97~\%}$, and~${\sim90.08~\%}$, respectively. All OCs with low reflectivities~${<2~\%}$ were measured in reflection and calculated with equation~\cref{eq:2}.

Prior to comparing the performance of the fiber laser with all OCs, we measured the center wavelength of the fiber laser with a flat cleaved OC. This measurement indicates the location of the highest gain within the bandwidth of the high reflecting mirror. We applied a precise strain to align the OCs center wavelength with the highest gain center wavelength. The accuracy of the alignment was within~${\pm0.1~nm}$ of the highest gain center wavelength~${\sim1546.2~nm}$. Each OC was spliced directly to the active fiber through a pump dump (the splice had a high index recoat material) and the output power was monitored with a constant pump power of~${\sim18.27~W}$. Total length of the fiber laser was~${\sim10~m}$ and output power was measured with a Coherent PowerMax PM30 power meter. Assuming a homogeneously broadened CW laser model with uniform distributed loss~\cite{rigrod1978homogeneously,schindler1980optimum}, we fit~(Fig.~\hyperref[fig:3]{\ref*{fig:3}}) the output power and OC reflection measurements with the following implicit equation:

\begin{figure}[!b]
\centering
\includegraphics[width=\linewidth]{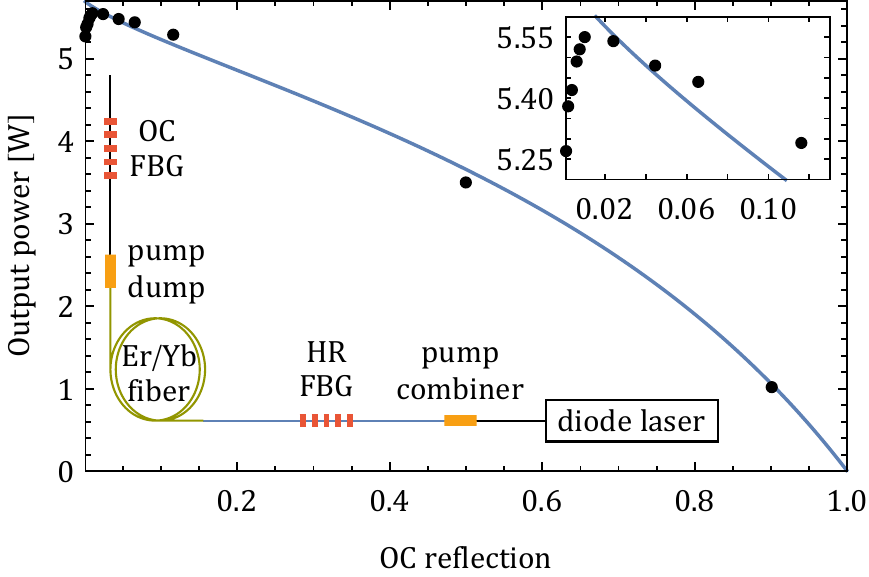}
\captionsetup{justification=justified}
\caption{Output power measurements with the OCs for a pump power of~${\sim18.27~W}$ and the model fit (blue curve). Figure inset shows the divergence of the model from the measurements at low reflecting OCs.}
\label{fig:3}
\end{figure}

\begin{equation}
\label{eq:3}
\dfrac{2g}{\zeta}arctan\left(\dfrac{AI_{s}(\alpha-g)(1-R)-2P\alpha\sqrt{R}}{AI_{s}(1+\sqrt{R})^{2}\zeta}\right)+ln(\sqrt{R})-\alpha l=0
\end{equation}

\noindent
where

\begin{equation}
\label{eq:4}
\zeta(g,\alpha,A,I_{s},P,R)=\sqrt{2g\alpha-g^{2}+\left(\dfrac{4P^{2}R}{A_{\phantom{s}}^{2}I_{s}^{\,2}(1-R)_{\phantom{s}}^{2}}-1\right)\alpha^{2}}
\end{equation}


\noindent
where~${g}$ is the unsaturated small signal gain, ${\alpha}$ is the loss coefficient per unit length, ${A}$ is the laser cross section, ${I_{s}}$ is the saturation intensity, ${l}$ is the gain material length and~${P}$ is the output power. We approximate the value of~${AI_{s}\sim0.0021~W}$ with wavelength, mode field diameter~(MFD), emission cross section and spontaneous decay time of~${\lambda=1546.2~nm}$, ${MFD=\SI{10}{\micro\metre}}$, ${\sigma=3\cdot10^{-21}~cm^{2}}$ and~${\tau=8~ms}$, respectively. Highest measured output power was~${\sim5.55~W}$ corresponding to an OC with a~${\sim1.01~\%}$ reflectivity. As evident from~Fig.~\hyperref[fig:3]{\ref*{fig:3}}, the model is in good agreement with the measurements for OCs with reflectivities higher than~${\sim3~\%}$; yet, as seen in the figure inset, the model fails to predict the optimal OC reflection, which is in the low reflectivity regime. The model curve diverges from our output power measurements for OC reflections smaller than~${\sim3~\%}$. As the OC reflectivity decreases, the model predicts a small increase in output power to~${\sim5.69~W}$ and then decreases rapidly to zero. According to our measurements, a substantial reduction of output power occurs for OCs with reflectivities higher than~${\sim10~\%}$.

In order to compare the optimal narrowband OC laser performance with a broadband flat cleaved OC performance, we measured the spectrum and output power of the fiber laser with two different broadband configurations: (a) front facet of the active fiber was flat cleaved, (b) front facet of the~\mbox{SMF-28} fiber was flat cleaved, rear facet was spliced directly to the active fiber through a pump dump.

\begin{figure}[!b]
\centering
\includegraphics[width=\linewidth]{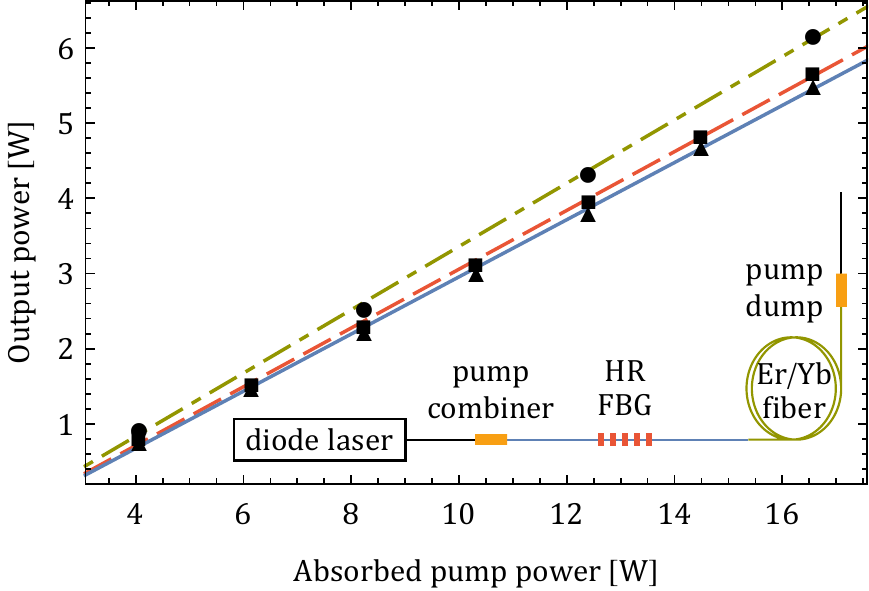}
\captionsetup{justification=justified}
\caption{Output power measurements with the broadband active (green dotted-dashed curve, circles), and~\mbox{SMF-28} (red dashed curve, squares) flat cleaved OCs. Blue curve (triangles) is the output power with the optimal narrowband OC with~${\sim1.01~\%}$ reflectivity. Output power measurements were performed for pump powers as high as~${\sim18.27~W}$. Figure inset shows the fiber laser layouts with the broadband flat cleaved~\mbox{SMF-28} OC.}
\label{fig:4}
\end{figure}

In~Fig.~\hyperref[fig:4]{\ref*{fig:4}} we show the efficiency of the fiber laser with the two broadband configurations and with the optimal narrowband OC. Efficiency of the fiber laser with a flat cleaved active fiber and flat cleaved~\mbox{SMF-28} was~${\sim42~\%}$ and~${\sim39~\%}$, respectively. With the narrowband OC, the fiber laser efficiency was~${\sim38~\%}$, which is very high considering the fact that the OC is inscribed on an~\mbox{SMF-28} and spliced directly to the active fiber. Surprisingly, similar output powers were obtained with the broadband flat cleaved and optimal narrowband~\mbox{SMF-28} OCs even though they have different reflectivities. One reason the broadband flat cleaved OC requires a higher reflecting OC compared with the optimal narrowband OC may have to do with longitudinal modes. With the broadband flat cleaved OC more longitudinal modes are competing on the same gain, thus requiring a higher reflecting OC to obtain a similar output power. Laser efficiency measurements with the optimal OC had an angle cleaved fiber facet that reduced additional reflections. The angle cleave was performed with a Vytran LDC-200, cleave angle was~${8\pm1\degree}$. We compared the stability of the output power with the broadband flat cleaved~\mbox{SMF-28} and with the narrowband optimal OC by monitoring the output power for~${\sim6~min}$. Largest power deviation for the broadband and narrowband configurations was~${<\pm0.86~\%}$ and~${<\pm0.85~\%}$, respectively.

\begin{figure}[!t]
\centering
\includegraphics[width=\linewidth]{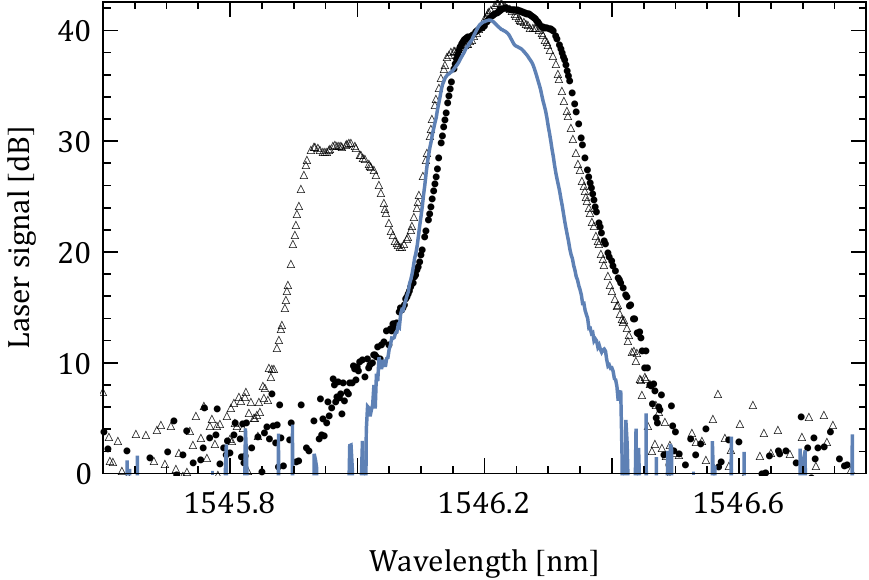}
\captionsetup{justification=justified}
\caption{Laser signal with a broadband flat cleaved~\mbox{SMF-28} OC (black circles) and narrowband OC (blue curve) with~${\sim1.01~\%}$ reflectivity. With the broadband flat cleaved~\mbox{SMF-28} OC, there were short instances where the spectrum was similar to the narrowband optimal OC spectrum (black triangles). Pump power in all measurements was~${\sim18.27~W}$.}
\label{fig:5}
\end{figure}

The spectrum of the fiber laser was monitored with an OSA for the broadband and narrowband configurations~(Fig.~\hyperref[fig:5]{\ref*{fig:5}}). During most of the measurement duration, the flat cleaved~\mbox{SMF-28} OC had a deformed spectrum. Since there is no FBG, the laser broadband spectrum is mainly affected by: (a) splice conditions, (b) the high reflecting mirror profile and (c) the gain profile within the high reflecting mirror bandwidth. In several short instances, we observed with the broadband OC a spectrum that was slightly more broadband than the narrowband optimal OC uniform spectrum. Output power and spectrum measurements demonstrate that the best approach to extract the highest output power is by flat cleaving the active fiber facet (and filtering out the remaining pump signal) or by flat cleaving the~\mbox{SMF-28} fiber. These approaches are not very practical since any minor damage to the flat cleaved facet will have a detrimental effect on the output power and spectrum. Splicing a narrowband~\mbox{SMF-28} OC directly to the active fiber slightly reduces the efficiency of the fiber laser; yet, the spectrum is improved. Inscribing the OC on a passive double-clad fiber will probably help increase the optimal output power due to better mode matching between the active and passive fibers. Yet, inscribing the OC on an~\mbox{SMF-28} fiber is a cost effective approach that optimizes the fiber laser output power while still maintaining a uniform and narrow output spectrum.

\begin{figure}[!t]
\centering
\includegraphics[width=\linewidth]{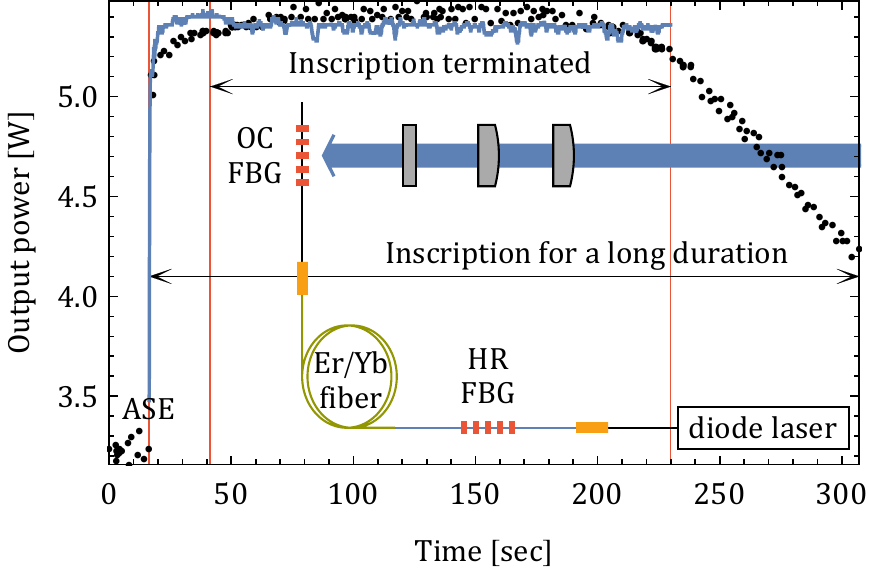}
\captionsetup{justification=justified}
\caption{Black circles are the output power measurements without terminating the OC inscription at the optimal output power. Blue curve is a separate measurement (with a different~\mbox{SMF-28} fiber) of the output power, where the OC inscription was terminated when maximum output power was achieved. An average output power of~${\sim5.35~W}$ was obtained with a pump power of~${\sim18.27~W}$.}
\label{fig:6}
\end{figure}

Optimization during laser operation is challenging since longer exposure times are known to red shift the center wavelength of the inscribed FBG~\cite{anderson1993production}; therefore, it was unclear if a similar output power to the optimal narrowband OC could be obtained by inscribing the OC in-situ during laser operation. This red shift can cause a small misalignment between the OC center wavelength and the highest gain center wavelength. In order to check if the fiber laser could be optimized in-situ we used the setup shown in the inset of~Fig.~\hyperref[fig:6]{\ref*{fig:6}}. As inscription time passes, the nonlinear absorption induces a larger refractive index modulation, and the OC reflection increases. Each OC inscription was repeated twice for a pump power of~${\sim18.27~W}$. The duration of the first inscription was much longer than the time necessary to optimize the output power. After reaching the highest output power we observed a slow decrease of output power over time (black circles in~Fig.~\hyperref[fig:6]{\ref*{fig:6}}) due to the increase in OC reflectivity. The second inscription was terminated when the highest output power was reached by blocking the femtosecond beam. The measured reflectivity of the OC inscribed in-situ was~${\sim0.20~\%}$, which explains why the measured average power of~${\sim5.35~W}$ was less than the optimal output power with a narrowband OC. Higher output powers can be obtained in-situ by improving the alignment between the inscribed OC center wavelength and the highest gain wavelength within the high reflecting mirror bandwidth. The spectrum of the fiber laser with the in-situ OC was similar to the spectrum with the optimal narrowband OC shown in~Fig.~\hyperref[fig:5]{\ref*{fig:5}}.

In conclusion, we inscribed twelve narrowband OCs on~\mbox{SMF-28} fibers and measured the optimal OC of a~${\sim10~m}$ Er/Yb fiber laser. After splicing the optimal OC, which had a~${\sim1.01~\%}$ reflectivity, directly to the active fiber we obtained an output power of~${\sim5.55~W}$ for a pump power of~${\sim18.27~W}$. We compared the output power and spectrum of the optimal narrowband~\mbox{SMF-28} OC to a broadband flat cleaved~\mbox{SMF-28} OC. Output power was slightly higher with a broadband flat cleaved OC; yet, the spectrum was more uniform and narrow with the optimal narrowband OC. We inscribed an OC in-situ with a similar reflectivity to the optimal narrowband OC, and demonstrated that a similar output power can be obtained with a uniform and narrow spectrum.

\noindent
\textbf{\large Funding.}\indent Israel Ministry of Industry, Trade and Labor, ALTIA Magnet program (60882); Israel Ministry of Science, Technology and Space (311877).


\bibliographystyle{osajnlnt}
\bibliography{refs}


\bibliographyfullrefs{fullrefs}

\end{document}